\documentclass[namedreferences,hyperref,optionalrh]{spr-sola}
\usepackage{graphicx}        % For eps figures, newer & more powerfull
\usepackage{color}           % For color text: \color command
\chardef\us=`\_

\begin{document}

\begin{frontmatter}
\title{Gnevyshev gap in the large-scale magnetic field}

\author[addressref={aff1},email={e-mail. obridko@izmiran.ru}]{\inits{V.N.}\fnm{V.N.}~\snm{Obridko}\orcid{ 0000-0001-5100-806X}}
\author[addressref=aff1, email={e-mail.as.shibalova@physics.msu.ru}]{\inits{A.S.}\fnm{A.S.}~\snm{Shibalova}\orcid{0000-0002-7228-6226}}
\author[addressref={aff2}, corref, email={e-mail.sokoloff.dd@gmail.com}]{\inits{D.D.}\fnm{D.D.}~\snm{Sokoloff}\orcid{0000-0002-3441-0863}}
%\author{\inits{}\fnm{}~\lnm{}\orcid{}}
%   NOTE:  Just one corresponding author [corref]
\address[id=aff1]{IZMIRAN, 4, Kaluzhskoe Shosse, Troitsk, Moscow, 142190, Russia}
\address[id=aff2]{IZMIRAN, 4, Kaluzhskoe Shosse, Troitsk, Moscow, 142190, Russia, Department of Physics, Lomonosov Moscow State University, Moscow, 119991, Russia and Moscow Center for Fundamental and Applied Mathematics, Moscow, 119991, Russia}

\runningauthor{Obridko et al.}
\runningtitle{Gnevyshev gap in the large-scale magnetic field}

\begin{abstract}
The phenomenon of the Gnevyshev gap was first identified in the solar-corona irradiance data (green line). Later, it was studied in the sunspot, coronal, and heliospheric data. We have investigated the Gnevyshev gap in the magnetic field data and have arrived at the conclusion that it reflects the behavior of the large-scale magnetic field. The Gnevyshev gap occurs at the 
polarity reversal of the solar magnetic field at the photosphere level. The presence of the Gnevyshev gap in sunspot data at the photosphere level is  disguised by non-global  structures that retain dependence on both latitude and longitude (the accepted mathematical term is tessaral, see below for more details). However, it is clearly visible in the magnetic field data at the photosphere level and is even more pronounced at the source surface (i.e., in the corona). 
%\texttt{sola\us keyword\us list.txt}.  
\end{abstract}
\end{frontmatter}
%-------------------------------------------------

\section{Introduction}
     \label{S-Introduction} 
In 1967, M.Gnevyshev \citep{G67} drew attention to the fact that, in the activity Cycle 19 (1954--1965),  the emission of the solar corona in the spectral line $\lambda = 5303$ \AA (green line) had two maxima --- in 1957 and in 1959-1960. A decrease in the half-year mean coronal brightness was called the Gnevyshev gap (GG). This phenomenon has since been observed in various indices, mainly related to the occurrence rate of solar chromospheric flares, the upper layers of the solar atmosphere, the interplanetary medium and cosmic rays. Some authors associated it with quasi-biennial variations of solar activity (for review, see  \citealp{Setal03, Betal06, Betal14}  and references therein, \citealp{Ketal15}).

It should be noted that the two-peak structure of the solar cycle is quite rare when the standard 13-month smoothing is applied (Fig.1). An indication of their existence can be seen only in 7 cycles (5, 11, 12, 16, 22, 23, and 24) out of 24 cycles for which reliable data are available (Fig. 1). However, in cycles 11. 12. 16, generally speaking, there are no dips; there are individual weak intensifications during the growth or decline phase, and the cycle remains single-peaked. In cycles 5, 22, 23, and 24, the dip is actually observed to be 10-15\% deep. The peak ratio varies from 1.098 to 1.184, with the highest secondary maximum observed in cycle 24, and only in cycle 22 is the secondary maximum lower than the first (see http://www.solen.info/solar/cycles1\_to\_present.html or https://www.sidc.be/SILSO/datafiles).

On a shorter timescale (monthly smoothing), we usually get  more than three peaks unevenly distributed over a wide range of about $12\pm6$ months. \cite{K05} suggested calling these structures the Gnevyshev peaks and Gnevyshev gaps. They can be identified in various photospheric indices and are even more pronounced in indices related to higher layers of the solar atmosphere (chromosphere and corona). \cite{K05} analyzed various interplanetary indices, such as the solar wind density, velocity, and magnetic field, as well as some geophysical indices, to find out that not a single  peak obtained by monthly averaging could be identified for the entire set of indices under investigation. 

\begin{figure}    %%%%%%%%%%%%%%%%%% FIGURE 1 
\includegraphics[width=0.49\textwidth]{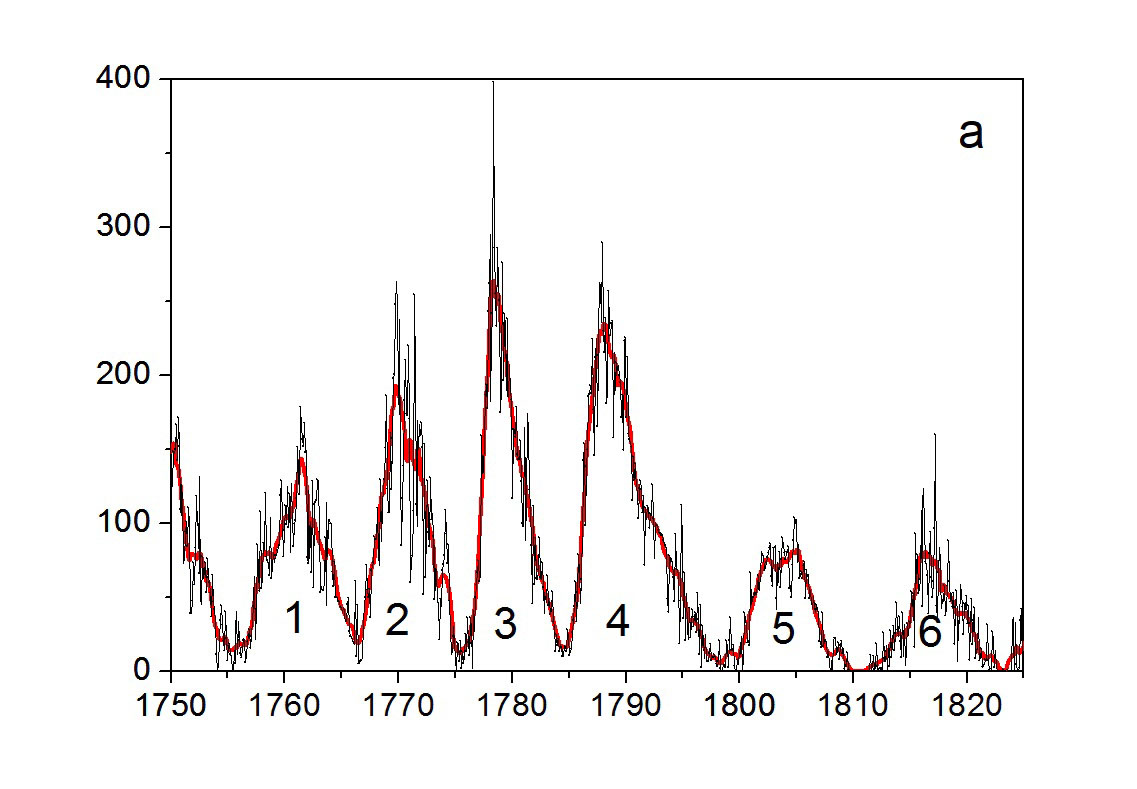}
\includegraphics[width=0.49\textwidth]{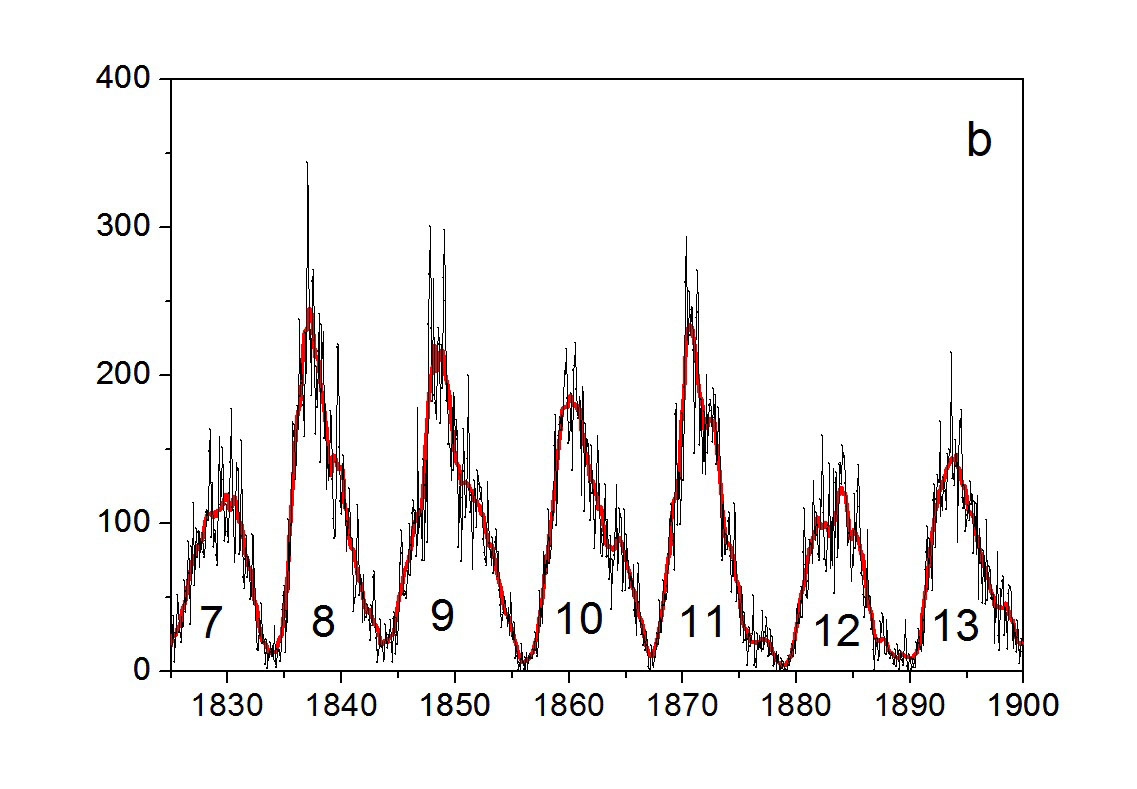}\\
\includegraphics[width=0.49\textwidth]{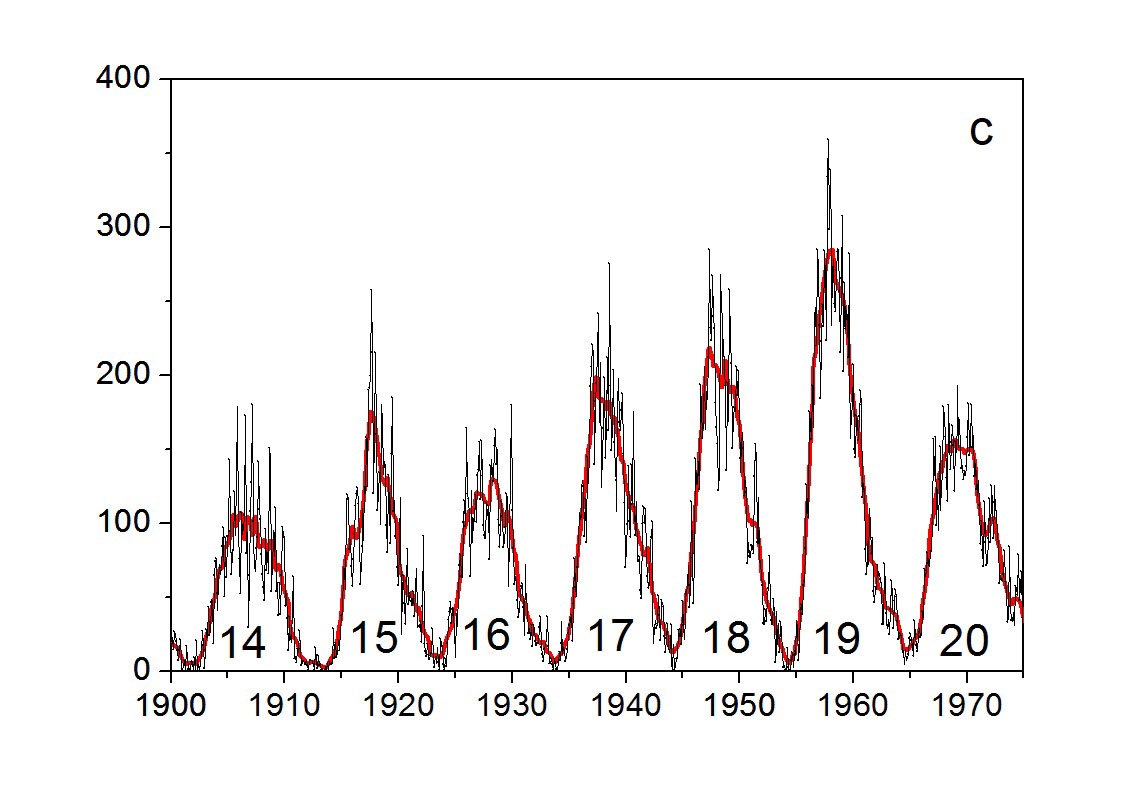}
\includegraphics[width=0.49\textwidth]{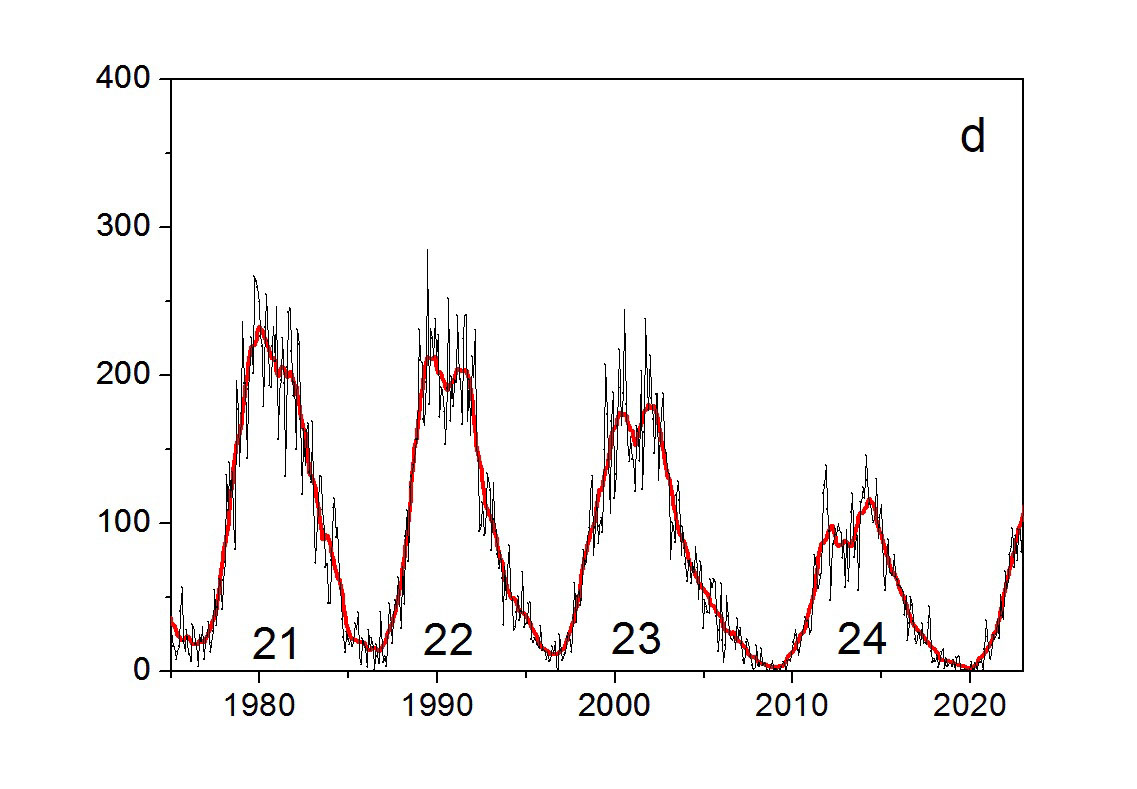}
\small
        \caption{ The dependence of sunspot numbers in Cycles 1-24 (divided into four panels) on time (years). The monthly mean numbers (vertical axis) are given in black; red shows the standard 13-month averaging. Time in years is given in horizontal axis. (Sunspot data from the World Data Center SILSO, Royal Observatory of Belgium, Brussels. Version V2)
                }
\label{F1}
\end{figure}

On the other hand, in the cosmic ray and interplanetary data, one can clearly see a time interval when the solar activity decreases. For this period we retain the name Gnevyshev gap (GG), although this effect may not actually be observed in the photosphere. 

Thus, the Gnevyshev gap concept looks  different in different solar activity tracers, and the overall situation looks quite complicated. It seems (cf. \citealp{Setal03}) as if there is a filter that prevents the GG-effect from propagating to higher levels of the solar atmosphere and into the interplanetary medium. On the other hand, \cite{Betal98} showed that the dipole-like component, which becomes minimal during GG, is not dominant in the photosphere, but its role increases in the corona.  

In our opinion, the GG concept needs clarification and a deeper physical interpretation. \cite{Betal98} pointed out a very important fact. So far, GG has been studied using sunspot data or data from high levels of the solar atmosphere or even the interplanetary medium, while the role of the global magnetic field has been outside our scope of attention. 
We believe that the data used so far without taking into account global fields are insufficient to understand the nature of GG. Indeed, it is reasonable to believe that GG may be associated with processes that happens in the solar interior or just at the solar surface. Therefore the surface data should be most informative for understanding the GG phenomenon (unfortunately, we can not observe the magnetic field in the solar interior directly). Sunspots are helpful in this regard; however, they occupy only a tiny part of the solar surface --- no more than a few thousandths even during the solar maximum. As for the magnetic flux, the sunspot contribution is also much smaller than the integral solar magnetic flux \citep{Oetal22}. 

 Here, it is natural to proceed from the well-known paper by \cite{WS03}, who described the GG phenomenon in terms of the magnetic-field transport by emerging sunspots. We appreciate the scenario presented therein, but our purpose is to expand the observational basis and involve in  consideration source-surface magnetic field.

We believe that to understand the nature of the Gnevyshev Gap it is necessary to involve data of the surface large-scale magnetic field, and this is the goal of our article. In particular, we consider the energy indices of the solar surface magnetic field to study their cyclic variations and clarify the nature of the GG phenomenon.

\section{Gnevyshev gap and energy indices of the solar large-scale magnetic field }

Let us apply the method of spherical analysis to estimate the contribution of different structural parameters to the cyclic evolution of large-scale fields.
We shall use photospheric magnetic field data obtained at Stanford with 3 arc min resolution for the period of June 1976 to July 2022, i.e., for Carrington rotations  1642—2055 obtained from http://wso.stanford.edu/forms/prsyn.html. We assume that in a spherical layer from the photosphere surface of radius $R_0$ to a fixed spherical surface of radius $R_s$, conventionally called the source surface, the magnetic field is completely described by the potential approximation PFSS \citep{Hoeks84, Hoeks91}. This yields the following set of magnetic field components $B_r, B_\theta, B_\phi$ in a spherical coordinate system $r, \theta, \phi$ (polar radius, latitude, and longitude, respectively):

 \begin{eqnarray}
 B_r=\sum_{l,m}P_l^m (\cos \theta) (g_i^m \cos \phi + h_i^m \sin m \phi)\times\\
 \nonumber
 \times ((l+1) (R_0/r)^{l+1} - l(r/R_s)^{l+1}c_l\,,\\
 B_\theta =-\sum_{l,mm}\frac{\partial P_i^m(\cos \theta)}{\partial \theta}(g_l^m \cos m\phi +h_l^m\sin m\phi) \times \\
 \nonumber
 \times ((R_0/r)^{l+2}+(r/R_s)^{l-1}c_l)\,,\\
 B_\phi = -\sum_{l,m} \frac{m}{\sin \theta} P_l^m (\cos \theta) (h_i^m \cos m \phi - g_l^m \sin m \phi) \times\\
 \nonumber
 \times ((R_0/r)^{l+2} +(r/R_s)^{l-1}c_l)\,.
\end{eqnarray}
 
Here, $0 \le m$, $l < N$ (usually, $N \le 9$), $c_l=-(R_0/R_s)^{l+2}$, $P_l^m$  are the Legendre polynomials, and $g_l^m$, $h_l^m$  are the harmonic coefficients.
The latter was calculated from WSO Stanford data. To find the harmonic coefficients, $g_l^m$ and $h_l^m$, and thus, to fully determine the solution, we had to use boundary conditions. The radial component of the magnetic field  on the photosphere surface plays the role of one boundary condition, while the other is the assumption  that the magnetic field at the source surface is purely radial.

In order to estimate the magnetic energy at the sphere of radius $r$ we introduce as an index the mean value of $B_r^2$ taken at the sphere of radius $r$ \citep{OE89, OE92}:

\begin{equation}
i(B_r)_{|r} = <B_r>_{|r}\,,
\end{equation}
which yields 
\begin{equation}
i(B_r)_{|R_0}=\sum_{l,m}\frac{l+1+l \zeta^{2l+1})^2}{2l+1} ((g_l^m)^2 + (h_l^m)2)
\end{equation}
and
\begin{equation}
i{B_r}_{|R_s}= \sum_{l,m} (2l+1) \zeta^{2l+4}((g_l^m)^2 + (h_l^m)2)
\end{equation}
for the surface and source radii, respectively. Here, $\zeta = R_0/R_s$. In our calculations, the radius of the source surface is $R_s = 2.5 R_0$ 
and, thus, $\zeta = 0.4$.

\begin{figure}    %%%%%%%%%%%%%%%%%% FIGURE 2
\includegraphics[width=0.90\textwidth]{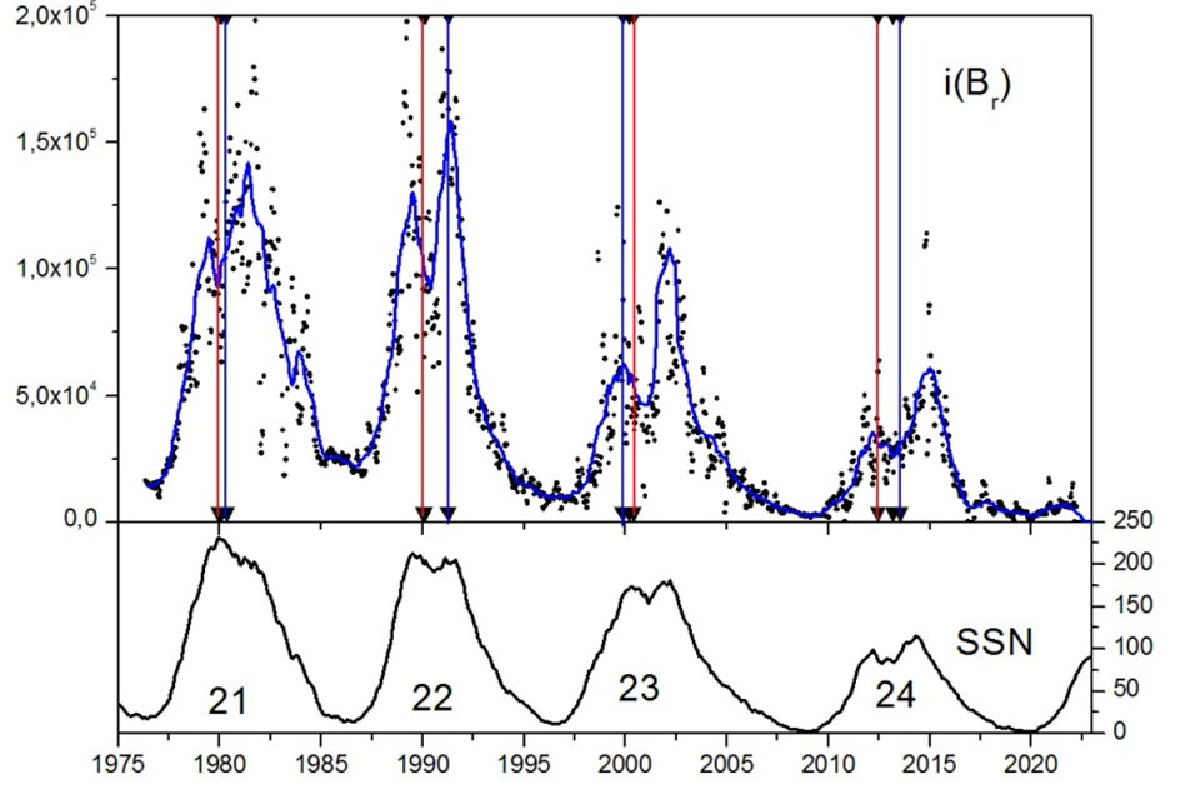}
\small
        \caption{Evolution of $i(B_r)$ at the photoshere for 1976--2022 measured in $\mu T^2= 10^{-4}$ G$^2$ (upper panel, black points show the values for each Carrington rotation, blue points show the data smoothed over 13 rotations) compared to sunspot data smoothed over 13 months (lower panel, cycle numbers are indicated, SSN are given on the right). The vertical red lines show the dates of magnetic field reversal at the North pole; the blue lines stand for the reversals at the South pole. The times of the polar reversals were taken from http://wso.stanford.edu/Polar.html.}
\label{F2}
\end{figure}

 Fig.~\ref{F2} illustrates the evolution of $i(B_r)$ calculated at the photoshere for the time interval 1976--2022, when the required observational data are available. The two-peak structure is well-pronounced in all four cycles presented. The later peak is always higher than the earlier one. The quantitative characteristics of the two-peak structure and the Gnevyshev gap are given in Table~\ref{T1}. 

\begin{table}[b]
\caption{Gnevyshev gap parameters in the last four cycles according to the surface mean magnetic field data.  $N_1$, $T_2$ are the dates of the first and second maxima respectively, $h_1$, $h_2$ are their heights, and $h_{GG}$ is the height of the Gnevyshev gap.}
\begin{tabular}{|c|c|c|c|c|c|}
\hline
Cycle number & $T_1$, yrs & $T_2$, yrs & $T_2-T_1$, yrs & $h_2/h_1$ & $h_{GG}/h_1$ \cr 
\hline
21 & 1979.3 & 1981.3 & 2.0 & 1.25 & 0.848 \cr
\hline
22 & 1989.4 & 1991.3 & 1.9& 1.23 & 0.73 \cr
\hline
23 & 1999.9 & 2002.4 & 2.5 & 1.74 & 0.726 \cr
\hline
24 & 2012.2 & 2015.0 & 2.8 & 1.69 & 0.778 \cr
\hline
\end{tabular}
\label{T1}
\end{table}

\begin{figure}    %%%%%%%%%%%%%%%%%% FIGURE 3 
\includegraphics[width=0.90\textwidth]{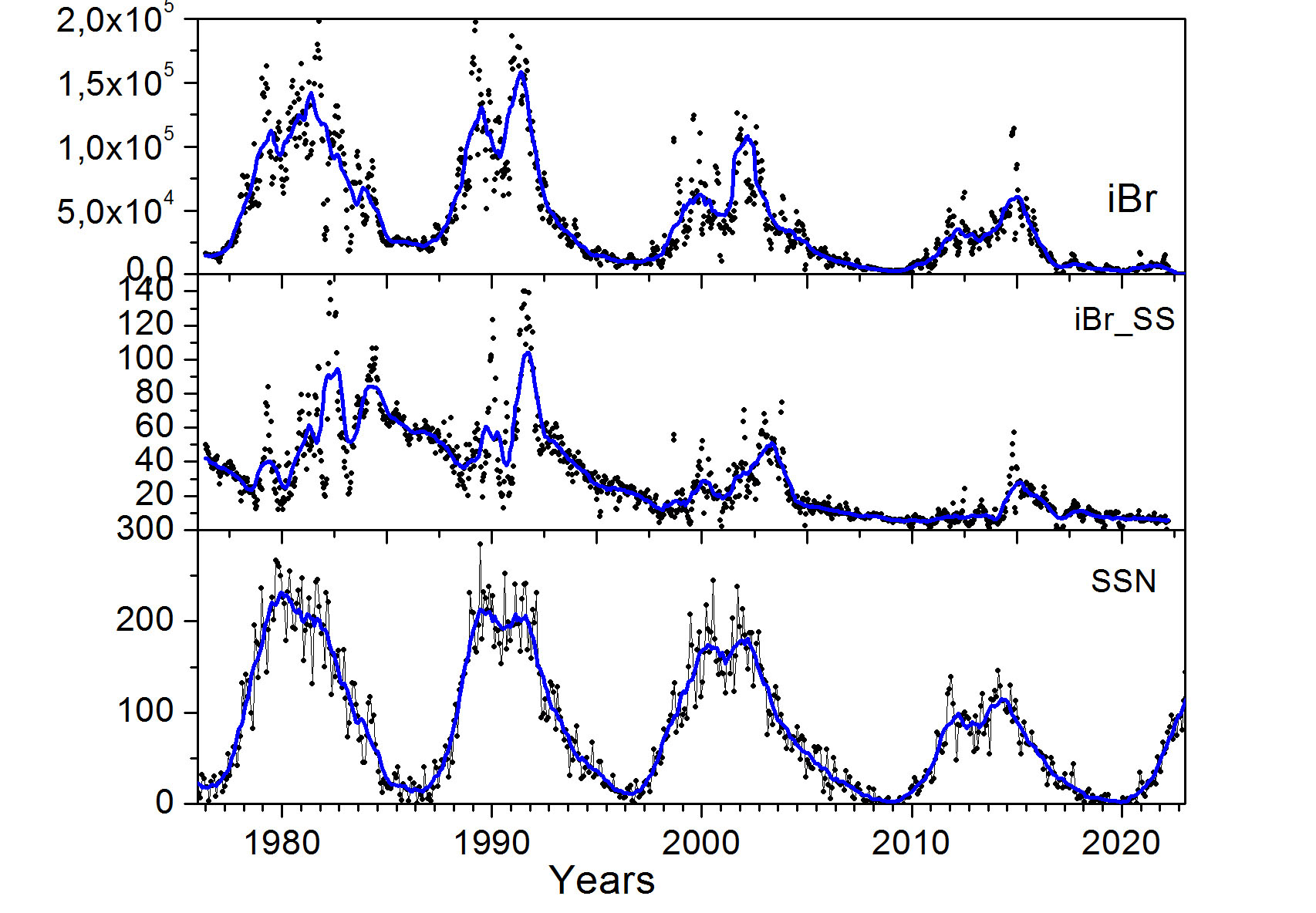}
\small
        \caption{Evolution of the energy index calculated at the photosphere level (upper panel) compared to that calculated at the source surface (middle panel), measured in $\mu T^2= 10^{-4}$ G$^2$ . The lower panel represents the variation of sunspot numbers.The black dots indicate the values for each Carrington rotation, the blue line shows the smoothing over 13 rotations.}
\label{F3}
\end{figure}

As a matter of fact, the phenomenon of the Gnevyshev gap actually consists of two related, but not identical processes (cf. \citealp{K05}) --- a moderate decrease in the cycle amplitude after the first maximum and the development of a secondary maximum. This becomes clearer if we compare the energy indices at the photosphere with those at the source surface, which are associated with  magnetic fields of larger scale, Fig.~\ref{F3}. The relative contribution of both phenomena depends on the level at which they are considered. 

As  seen in Fig.~\ref{F3}, the Gnevyshev gap is more pronounced at the source surface  than at the photosphere. More precisely, the first peak at the source surface is significantly lower than at the photosphere, and the second one is significantly higher.

\section{Magnetic field symmetries and the structure of the Gnevyshev gap.}

The processes associated with the GG phenomenon can be described in more detail by considering the contribution of magnetic fields of different symmetries.

The large-scale magnetic fields can be divided by harmonic coefficients into the following structures of different symmetries \citep{OE89, Setal89, OE92}:
\begin{itemize}
 \item Zonal structures ($m=0$). Harmonics are called zonal if their values change only with latitude. Within one zone, separated from others from the north and south by parallels, the zonal harmonic retains its sign. These, in turn, can be divided into zonal odd -- ZO ($l$ is odd, the most important case is the axial dipole, $l$=1) and zonal even --
ZE ($l$ is even, the most important case is the axial quadrupole, $l$=2);
 \item Sectorial structures $m=l$. Harmonics are called sectorial if their values change only with longitude. Within one sector, separated from others by meridians, the sectorial harmonic retains its sign. These, in turn, can be divided into sectorial odd -- SO ($l$ is odd, the most important cases are the equatorial dipole and two-sector structure with $l=m=1$) and sectorial even -- SE ($l$ is  even, say $l=m=2$);
 \item Tesseral structures ($m>l$); i.e., all other harmonics. These harmonics are so named because the curves on which they vanish are l-m parallels of latitude and 2m meridians, which divide the surface of a sphere into quadrangles whose angles are right angles (tessaras) \citep{Whitwat90}. 
\end{itemize}

\begin{figure}    %%%%%%%%%%%%%%%%%% FIGURE 4 
%\begin{minipage}[t]{0.60\textwidth}
\includegraphics[width=0.80\textwidth]{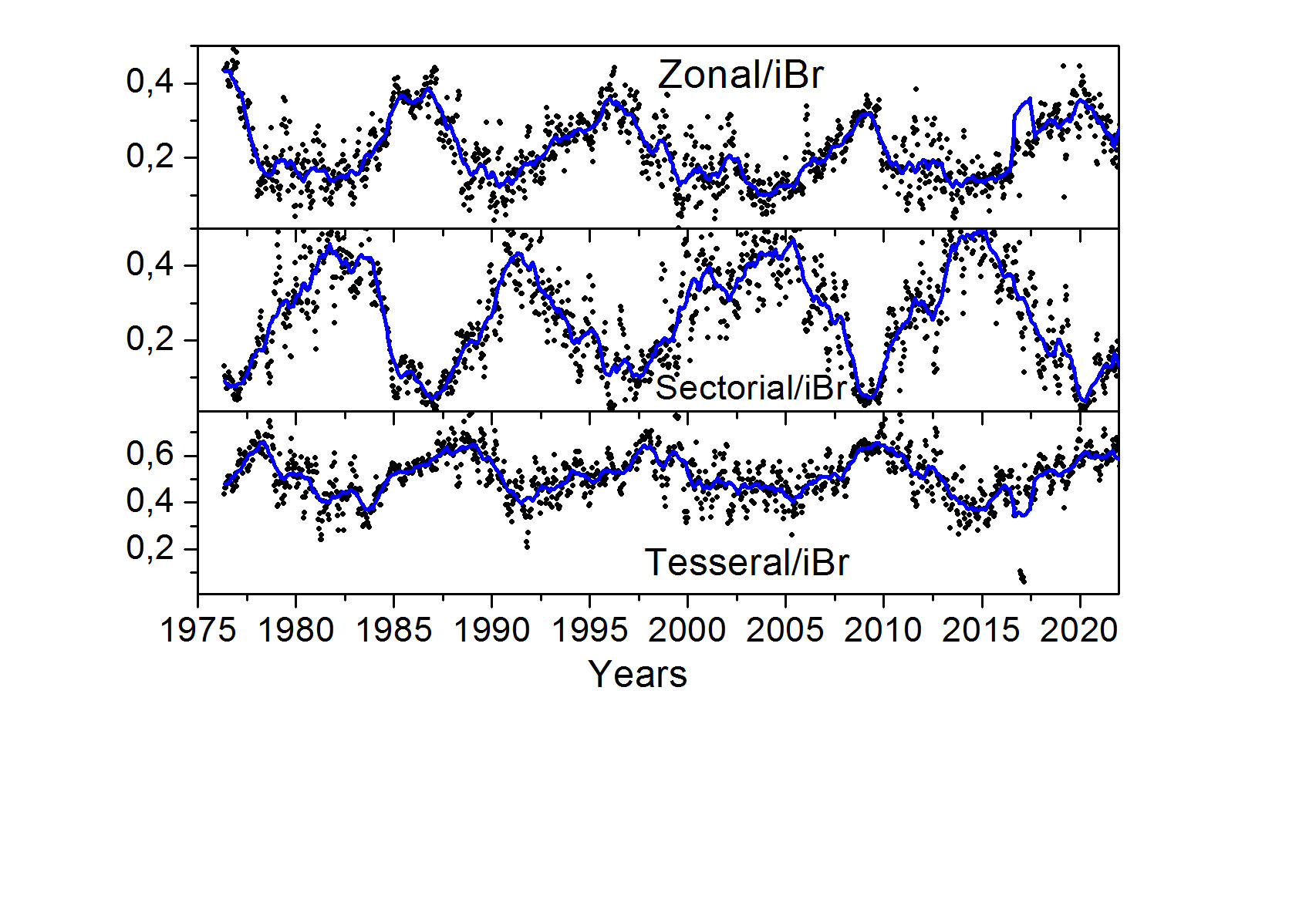}\\
%\end{minipage}
%\begin{minipage}[b]{0.60\textwidth}
\includegraphics[width=0.83\textwidth]{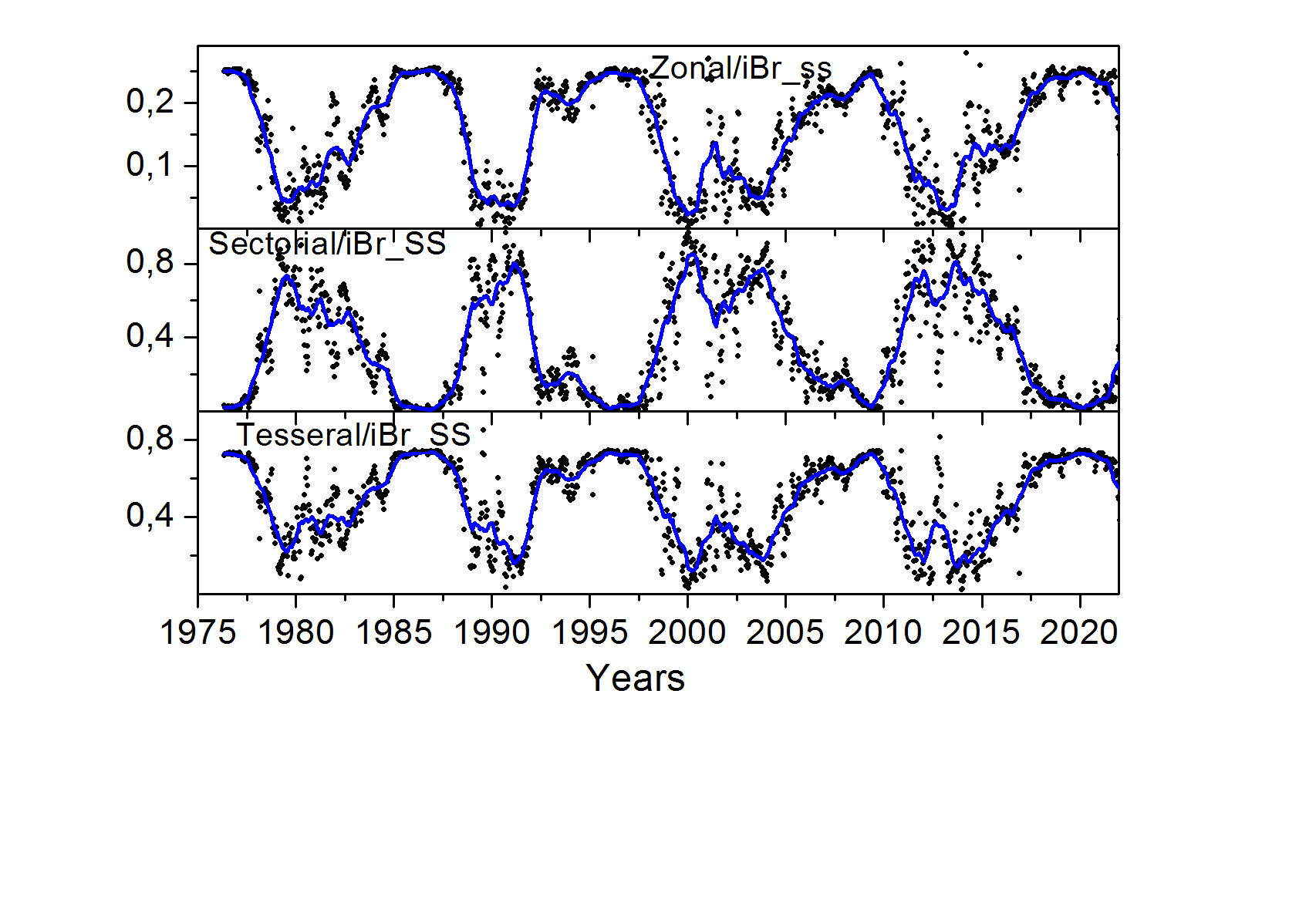}
%\end{minipage}
\small
        \caption{Evolution of energy indices at the photoshere  (top) and at the source surface (bottom). The upper row shows the contribution of the zonal harmonics  (measured in $\mu T^2= 10^{-4}$ G$^2$) , the middle row is the contribution of the sectorial harmonics, and the lower row corresponds to the tesseral harmonics. The indices are normalized to the total magnetic energy at the level under consideration.
                }
\label{F4}
\end{figure}

Of course, for each type of symmetry it is possible to introduce partial energy indices, the sum of which will be equal to the general indices discussed above. We performed a such calculation and normalized the partial indices to the general ones in order to exclude the general drop of the solar activity during the last solar cycles (Fig.~\ref{F4} left panel and Fig.~\ref{F4} right panel for the photosphere level and the source surface, respectively).  

Comparing Fig.~\ref{F3} and Fig.~\ref{F4} we see that the GG in the photoshere is associated with a substantial drop of the zonal harmonics. At the photosphere level the drop is slightly compensated since the tesseral harmonics associated with intermediate and small space doesn't change very much. The second peak is associated with the increase in sectorial harmonics, which occurs later.  The GG is more pronounced at the source surface where the relative contribution of  sectorial harmonics is much larger than that of the zonal and tesseral ones. This explains why the GG phenomenon associated mainly with the sectorial harmonics  was first isolated and then studied using coronal and heliospheric data.      

To summarize the above, we can conclude that the GG phenomenon is associated with the evolution of the large-scale magnetic field and arises due to the fact that different field components have a phase shift relative to each other. The gap occurs because the contribution of the zonal (i.e., axisymmetric) harmonic, especially its odd part (ZO),  decreases substantially. Later, at the end of the maximum phase, the contribution of the zonal even (ZE) and the sectorial harmonics increases. This is why the sectorial components of the interplanetary magnetic field are most pronounced at the beginning of the decay phase. Remarkably, the GG is most clearly seen in the indices where the contribution of the large-scale magnetic field is best pronounced. 

It is significant that the Z0 component associated with the rotation of the magnetic dipole is most pronounced during the reversals of the polar magnetic field. Note that the field reversals in the opposite solar hemispheres are not strictly simultaneous. So, there are time intervals when the magnetic field at both poles has the same sign. Such periods were the longest in Cycles 22 and 24 (see Fig.2 and Table 1).  
	
\section{Internal structure of the Gnevyshev gap}
More information about the internal structure of the Gnevyshev gap can be obtained from large-scale magnetic field data. Let us analyze them using Cycle 22 as an example.  A year and a half before the maximum (18 June 1988), we see at the source surface a standard picture dominated by the zonal structure. The north hemisphere is almost entirely occupied by the field of negative polarity and the south hemisphere, by the field of positive polarity. The situation changes dramatically in half a year (26 December 1988): the positive polarity penetrates  the north hemisphere while the negative polarity penetrates the south one. Even more changes are seen for 08 June 1989: the positive polarity has approached the north pole.  At the end of this year, the positive polarity reaches the north pole, which can be regarded as the polarity reversal. However, the positive polarity does not cover the whole north hemisphere, while the negative polarity has not yet reached the south pole. At this time, the sunspot number in the photosphere becomes maximum, which by convention is indicative of the the solar maximum. However, this second maximum in the sunspot data is lower than the first one, while in the photospheric magnetic field data, the second maximum is higher than the first one (see Fig.~\ref{F1}) (see Tabl.~\ref{T1}).

\begin{figure}    %%%%%%%%%%%%%%%%%% FIGURE 6 
\includegraphics[width=0.85\textwidth]{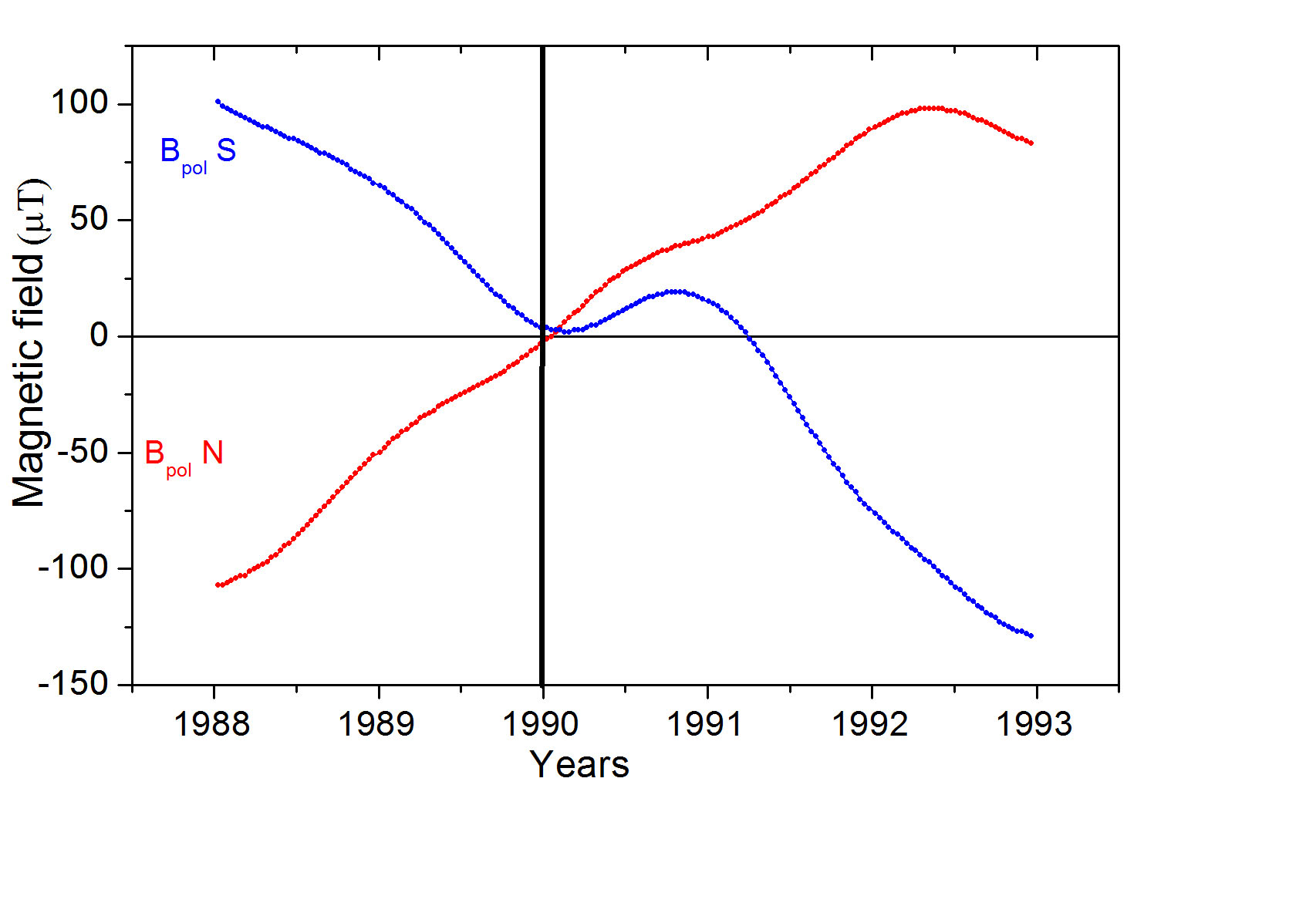}
\small
        \caption{Evolution of the polar magnetic field at the north (red curve) and south (blue curve) poles. The vertical line marks the minimum of the Gnevyshev gap.
       }
\label{F6}
\end{figure}

Later, in the first half of 1990, the sunspot number and the magnetic field decrease.  This is precisely what is called the Gnevyshev gap. The zonal odd  harmonics are virtually absent, and the field structure is similar to the equatorial dipole or quadrupole. The positive polarity dominates at the north pole until the next field reversal in Cycle 23 when the polar magnetic field (see Fig.~\ref{F6}) changes sign. In contrast, both polarities are visible near the south pole. This situation without a distinct prevalence of one or the other polarity lasts almost 2 years, and the field reversal in the south hemisphere takes place only in April 1991. At that time, the magnetic field increases substantially giving rise to the second maximum, which is most pronounced in the evolution of the source-surface magnetic field. The standard structure when each hemisphere is dominated by a single polarity and the magnetic field is determined by the zonal odd harmonic is only restored in the middle of 1992. 

To summarize, we can say that, the Gnevyshev Gap does not manifest amplitude variations in one or several magnetic-field tracers, but demonstrates a rather complex spatial interaction of its different components.

\section{Conclusion}

The above results lead us to the conclusion that the Gnevyshev gap reflects the behavior of the large-scale magnetic field. It occurs at polarity reversals of the magnetic field at the photosphere level. The presence of GG in sunspot data is disguised by non-global  (mainly tesseral) structures; however, it becomes clearly visible in the magnetic data in the photosphere and, especially, at the source surface (i.e., in the corona). That is why the GG was first identified in coronal observations and was later investigated using the indices associated with the magnetic field structure in the upper layers of the solar atmosphere or in the heliosphere. In general, the old dispute about the nature of the Gnevyshev gap (see  \citealp{Setal03, Betal06, Betal14}  and references therein, \citealp{Ketal15}) seems to be clarified.

Regarding the theoretical interpretation of GG by \cite{WS03}, we appreciate the scenario presented therein; however, we emphasize that the entire description of the phenomenon must explicitly include a large-scale magnetic field component. The theoretical ideas of \cite{WS03} are focused on the large-scale magnetic field and we present the observational basis of the GG using mean-field data.

According to the original definition (Gnevyshev, 1967), the GG phenomenon is commonly understood as a significant dip near the cycle maximum on the curve of sunspot numbers smoothed by the standard procedure (13-month smoothing) with two significant peaks before and after it. Smaller peaks and valleys can be observed in the monthly mean curve of sunspot numbers  during the cycle in all its phases. They can be caused by the emergence of one or several active regions or, conversely, by their absence and are not identified as GG phenomena.

The gap  is closely related to the phenomenon of reversal of the polar magnetic field. Of course, the polar magnetic-field reversal itself is not the initial cause of the gap, but rather serves as an indicator. Therefore, the gap can be observed somewhat later than the polarity reversal, as was the case in the Cycle 23. The immediate physical mechanism of the gap occurrence is the   changing relationship between the components of the large-scale magnetic field: a sharp decrease in the zonal and increase in the sectorial component, which leads to  discontinuity and the subsequent peak. On the photosphere surface, the gap is less pronounced, since the fields of smaller scales depend weakly on the cycle. In the corona, the contribution of small scales drops, and, therefore, the GG phenomenon is seen much more clearly. Following \cite{WS03} we stress that the polar magnetic field reversal happens together with the growth of equatorial dipole component which plays a key role in the GG formation at least during the recent cycles.

\begin{acks}
 The authors thank Todd Hoeksema for presentation of the data.
\end{acks}

\begin{authorcontribution}
All authors contributed to the study conception and design. Material preparation, data collection and analysis were performed by ASS under the guidance of VNO and DDS. The first draft of the manuscript was written by VNO and all authors commented on previous versions of the manuscript. All authors read and approved the final manuscript.
\end{authorcontribution}

\begin{fundinginformation}
VNO and DDS acknowledge the support of the Ministry of Science and Higher Education of the Russian Federation under the grant 075-15-2020-780 (VNO) and 075-15-2022-284 (DDS). DDS thanks support by BASIS fund number 21-1-1-4-1 and the financial support of the Ministry of
Education and Science of the Russian Federation as part of the program
of the Moscow Center for Fundamental and Applied Mathematics under the
agreement № 075-15-2022-284.
\end{fundinginformation}

\begin{dataavailability}
All data on photospheric magnetic fields are available at the\\ 
http://wso.stanford.edu/synopticl.html
And  Sunspot data from the World Data Center SILSO, Royal Observatory of Belgium, Brussels. Version V2. 
\end{dataavailability}

%\begin{materialsavailability}
%Information about available material ...
%\end{materialsavailability}

%\begin{codeavailability}
%Information about available code ...
%\end{codeavailability}

\begin{ethics}
\begin{conflict}
The authors declare that they have no conflict of interest.
\end{conflict}
\end{ethics}
  
\bibliographystyle{spr-mp-sola}
     % name your Bibtex file containing your references (.bib)
\bibliography{sola_bibliography_example}

\end{document}